# Metrology of Complex Refractive Index for Solids in the Terahertz Regime Using Frequency Domain Spectroscopy

Chick, S. R.[1,2], Murdin, B. N.[2], Matmon, G.[3] & Naftaly, M.[4]

[1] - Corresponding Author. Email: s.chick@surrey.ac.uk
[2] - Advanced Technology Institute, University of Surrey, Guildford, GU27XH, UK.
[3] - London Centre for Nanotechnology, University College London, London, WC1H 0AH, UK
[4] - National Physical Laboratory, Hampton Road, Teddington, Middlesex, TW11 0LW, UK

## Abstract

Frequency domain spectroscopy allows an experimenter to establish optical properties of solids in a wide frequency band including the technically challenging $10\,THz$ region, and in other bands enables metrological comparison between competing techniques. We advance a method for extracting the optical properties of high-index solids using only transmission-mode frequency domain spectroscopy of plane-parallel Fabry-Perot optical flats. We show that different data processing techniques yield different kinds of systematic error, and that some commonly used techniques have inherent systematic errors which are underappreciated. We use model datasets to cross-compare algorithms in isolation from experimental errors, and propose a new algorithm which has qualitatively different systematic errors to its competitors. We show that our proposal is more robust to experimental non-idealities such as noise or apodization, and extract the complex refractive index spectrum of crystalline silicon as a practical example. Finally, we advance the idea that algorithms are complementary rather than competitive, and should be used together as part of a toolbox for better metrology.

## Introduction

As the terahertz region above $3\,THz$ becomes more in demand for scientific & industrial applications, with new light sources [1] [2], nonlinear media [3], applications in both ultrafast science [4] and spectroscopy [5], and quantum technologies [6] the metrology of optical constants will be of elevated significance. Different experimental equipment can address different spectral regions, with vector- or scalar- network analysis appropriate for the $\sim\!100\,GHz$ range [7], Time Domain Spectroscopy (TDS) for the $\sim 3\,THz$ range [5], and ellipsometry across the spectrum, especially close to $\sim\!100\,THz$ [8] [9] all being independently studied, where differing considerations and terminology apply. In addition, recent experiments in terahertz travelling standards and instrumental intercomparisons [10] have highlighted the utility of Fourier Transform Infrared spectroscopy (FTIR) techniques as a calibration cross-check with sufficient bandwidth to be an effective comparison for all other methodologies. Continual improvement of the metrology surrounding FTIR techniques is therefore of direct relevance to the general effort of optical parameter characterization in the terahertz regime.

FTIR, as a means to perform c.w. Frequency Domain Spectroscopy (FDS), retains significance due to its ubiquity and relative cost efficiency, as well as decades of instrumental improvement, but especially because it can reach the $10 - 30\ THz$ region which all other techniques find challenging [11] [12] [13]. Experimental FDS strategies for obtaining refractive index (RI) of solids must take into account whether the sample is plane-parallel, in which case Fabry Perot (FP) interference modifies the sample's optical properties. While some techniques achieve high accuracy using multiple spectra of the sample – in reflection and transmission, for example [14] – to obtain sufficient information about the material, others leverage the form of FP interference to minimize the demands upon experimental data. The ubiquity of the apparatus lends itself to simpler methods which require only one measurement (plus a reference) to be made, so we focus in this paper upon techniques which can be implemented using a commercial transmission-mode FTIR machine with a broad-bandwidth source.

Historical literature for extracting optical constants from the Fabry Perot interference in FDS (FP-FDS) is plentiful [15] [16] [17] [18] [19] [20] [21] [22] [23] [24], but has ignored many of the systematic errors implicit in extraction techniques. Indeed, even comprehensive recent studies [10] [22] leave the sensitivity of their methods uncharacterized. Additionally, the reliance upon physical artefacts for intercomparison purposes leaves us comparatively ignorant about deviations of our measurements from the actual physical constants, either due to noise or systematic errors. Improved metrology for optical constants obtained through FTIR measurements is therefore an essential target for study and a relatively open field at the present.

This paper makes two advances: first we outline an improved methodology for extracting the real and imaginary components of the refractive index of a sample from its FP-FDS transmission function; and second we utilize well known models of that transmission function to characterize the algorithm. By using an analytical model of the transmission function, we obtain precisely the systematic errors implicit in the data analysis and compare quantitatively the sensitivity of different methodologies to simulated experimental non-idealities such as signal noise and frequency resolution. We show that our proposed algorithm overall outperforms an optimized version of more well-known methods, especially under experimental noise or limited resolution. We then implement a measurement of the complex refractive index of silicon in the band $2 - 19.5\ THz$ and use it to show that our algorithm is very robust to sources of incoherence in the measurement such as surface roughness, where simpler popular techniques are not.

Finally, we advocate that future experimenters have at hand a set of different tools for extracting optical constants from FTIR data – a sort of "toolbox" from which the appropriate algorithm can be chosen. Analytically modelling measured transmission functions using our methodology allows a metrologist to select the best tool for the job with increased confidence, thus refining the discourse on metrology in the terahertz regime.

## Theory

In this section, we explore the theoretical aspects of algorithmic refractive index extraction using the Fabry-Perot transmission function and its Fourier transform. We introduce common algorithms as a point of reference for our study, beginning with the commonly known fringe methods and advancing to Fourier based methods, discussing the systematic errors inherent in each. Our algorithm is introduced and discussed in comparison. The extraction of the imaginary part of the RI relies upon accurate knowledge of the real part so we discuss both components in detail.

Under linearly polarized illumination normal to plane-parallel surfaces, an etalon with complex refractive index $\hat{n} = n + ik$ and thickness $d$ has the FP-FDS transmission function $T(f)$ [25] [22] [18]:

1. $T(f) = \frac{\frac{1-\gamma^2 e^{-2\alpha d}R^2}{1-e^{-2\alpha d}R^2}\frac{(n^2+k^2)}{n^2}(1-R)^2\,e^{-\alpha d}}{1+\gamma^2 R^2 e^{-2\alpha d}-2\gamma Re^{-\alpha d}\cos[\Theta]}$

Where $c$ is the speed of light, $\alpha = 4k\pi/\lambda$ is the absorption coefficient. The phase $\Theta = 2\phi + 2nf\frac{2\pi}{c}d$ increases linearly with frequency and gives rise to the FP interference fringes. The "coherence fraction" $\gamma$ describes the relative coherence between successive internal reflections in the substrate [22]; $\gamma = 1$ represents total coherence, $\gamma = 0$ an entirely incoherent process. Unless otherwise stated, we assume in this paper that $\gamma = 1$. The surface reflectance of the substrate, $R$, and the phase shift on internal reflection due to absorption, $\phi$, are functions of $\hat{n}$:

2. $R = \left(\frac{(1-n)^2+k^2}{(1+n)^2+k^2}\right)$
3. $\tan\phi = \frac{2k}{n^2+k^2-1}$

We expect that an experimenter measures $T(f)$ and then wishes to obtain $\hat{n}(f)$ with maximum accuracy and using as little *a priori* information as possible. It is expected that standard data processing methods are available to the experimenter – fast Fourier transforms and curve fitting – but we would like to minimize the computational load if possible, thus prejudicing against techniques which rely very heavily upon local curve fitting to extract $\hat{n}$ explicitly.

Finally, we assume throughout that the experimental data are sufficiently well resolved that the fringes of Eqn. 1 are visible in the data. Where samples are optically thick, or the resolution of the spectrometer is too low, the fringes are insufficiently resolved to apply these algorithms. As we will see, some algorithms require thick samples and high resolution to measure accurately and with small datapoint spacing, thus they are more difficult to apply to a wide variety of different samples in a common lab setting.

## Divorcing Experimental and Computational Errors

Typically in RI extraction using FP-FDS, an algorithm is introduced and studied theoretically and then applied to a measurement of a relatively well-known sample such as silicon or quartz in frequency bands excluding sharp or strong absorption features. Occasional cross-comparisons between results are found [22], but it is often ambiguous whether any limitations might be fundamental to the algorithm or a feature of the experiment. Better metrology can be enabled by deliberately divorcing these sources of error so that the data analysis algorithms can be compared independently of data.

Therefore we will graphically compare the different algorithms by applying them to an artificial dataset based on Eqn. 1 and a model $\hat{n}$. Our model data use Eqn.s 1-3 to calculate a model transmission function $T_{model}(f)$ from an input $\hat{n}(f)$. The model $n$ & $k$ are generated by defining $k_{model}(f)$ over the frequency interval of interest, and using a published Kramers-Kronig (KK) transform library [26] to produce the corresponding variation in refractive index $\Delta n_{model}$. We then add a constant $n_{add}$ (since $n_{add}$ is not defined by the KK transform of $k$) so that we have a physically consistent complex refractive index of magnitude relevant to solid state physics, similar to that of silicon in the $2-20\,THz$ range. We chose $k_{model}(f)$ to be a sum of Gaussians, since they decay to zero rapidly which improves the accuracy of the numerical recipe for the KK transform. For example, we choose two Gaussians of different amplitudes and widths to approximate the observed absorption function in real data of Si etalons (for example, see Results section of this work). Figure 1 shows the corresponding $n$ with a value of $n_{add} = 3.4153$ and an example simulated FP interference spectrum result of Eqn. 1. Our model data can be replicated trivially using our code published under the GNU general license [29].

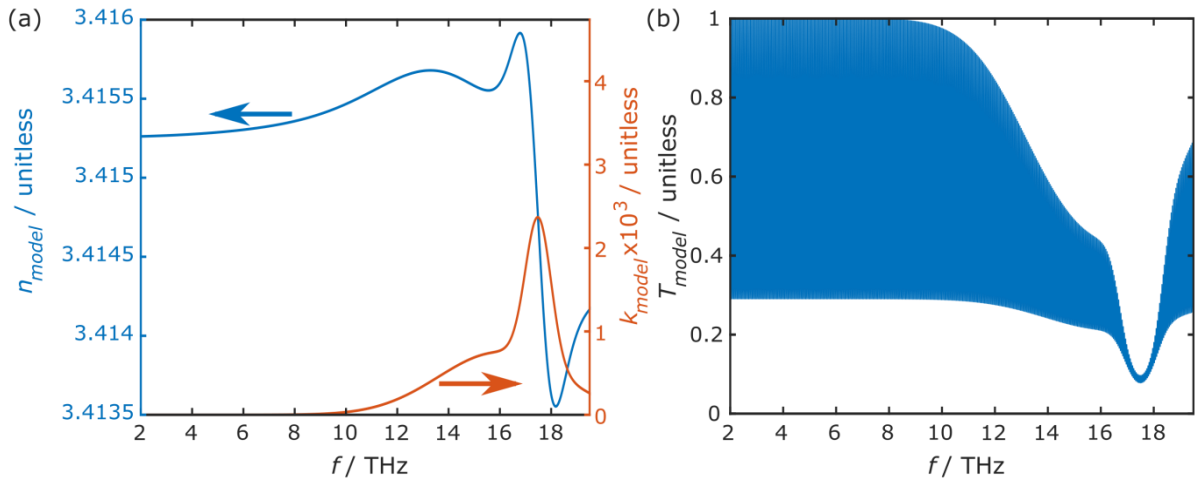

*Figure 1 – Model data used in the theoretical parts of this work. (a) Modelled complex refractive index derived by numerical KK transform; real part (blue) derived from the defined imaginary part (red). (b) Model transmission function $T_{model}$ obtained by substituting the data from (a) into Eqn. 1 with an etalon thickness of $d = 1\ mm$ and $\gamma = 1$.*

The dataset shown in Figure 1 will be used for cross-comparison of techniques throughout this Theory section. This does not eliminate the need for a careful experiment to validate our conclusions. Experimental data is invaluable in showing how our data analysis fails in real samples – and how it should be improved in future. We leave this effort to the Results section.

### Frequency Domain Fringe Methods

We now turn to practical methods of RI extraction. Our first point of discussion is the simplest set of methods by which we may extract $\hat{n}$. We shall relate the frequencies $f_\pm$ of local maxima ($T_+$) or minima ($T_-$) to the refractive index, from which a family of algorithms arises which we denote "fringe" methods [21] [10]. By inspection of the periodic term in the denominator of Eqn. 1, it is straightforward to show that $T_+$ and $T_-$ satisfy:

4. $\quad \phi + nf_- \frac{2\pi}{c} d = \pi\left(m + \frac{1}{2}\right), \phi + nf_+ \frac{2\pi}{c} d = \pi m$

The integer $m = 0,1,2 \dots$ is called the order of the fringe, and these relations imply that the real part of the refractive index at any point $f_\pm$ may be deduced if $m$ is known absolutely, and $\phi$ is known or negligible. If $m$ is unknown, it may be determined by locating each fringe in the dataset and extrapolating the observed fringe order to zero-frequency, which we term an "extrapolation method". However, extrapolation will only yield a correct result if the refractive index is constant over the range of data chosen for extrapolation. The order of the fringe may be over- or underestimated depending on the sign of the derivative of $n$ with respect to $f$. We demonstrate in Figure 2 how this error manifests as a systematic error when an extrapolation method is applied to $T_{model}$. Since the fringes are not equally spaced, the estimated $m$ is incorrect and a systematic error of form $1/f$ is observed. When the error in the extrapolated fringe order is less than one half, rounding $m$ can eliminate the systematic error, but since it is not in general easy to tell whether this will be the case, we choose not to round $m$ in this paper.

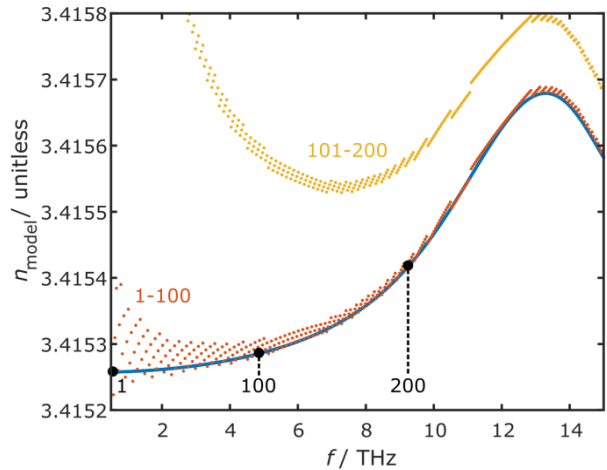

*Figure 2 – Systematic errors in refractive index extraction by fringe extrapolation methods. We use a peak finding algorithm to obtain $T_+$ from the model data of Figure 1 and extrapolate the fringe locations over two different regions to show how variation in the rate of change of $n_{model}$ affects subsequent extracted $n$. (blue) the reference dataset $n_{model}$. (red) extracted $n$ where the first 100 fringe maxima in $T_{model}$ were used for extrapolation to obtain the absolute fringe order, m. (yellow) extracted $n$ where the next 100 fringe maxima in $T_{model}$ were used for extrapolation to obtain the absolute fringe order, m. (black) the measured locations of the $1^{st}$, $100^{th}$, and $200^{th}$ fringe maxima in the model data $T_{model}$. Discretisation errors are also present in this analysis, manifesting as fine structure in the extracted $n$.*

If $m$ is unknown and the experimenter is not confident in extrapolation, the simplest approximation is to assume that the RI is sufficiently slowly varying that it is effectively constant for any two successive fringes, from which an expression may be found for the refractive index between the $m^{th}$ and $(m + 1)^{th}$ order fringes:

5. $\quad n[f_{(m+1)+} - f_{(m)+}] \frac{2\pi}{c} d = \pi$

This "fringe-difference" method yields correct results regardless of $\phi$ if the refractive index is constant, but when in fact $n$ varies substantially there is systematic error as illustrated later in this section. This method therefore shows a strong systematic error in any case where the transmission spectrum amplitude is not changing very slowly. Although the fringe-difference method is commonly used [21], its systematic errors are obviously related to this assumption of unvarying refractive index.

It is also important to extract the imaginary part of the refractive index. The effect on the transmission function of Eqn. 1 of $k(f)$ is obvious in the case Figure 1 where increased absorption causes a reduced fringe amplitude. There are several approaches to extracting this quantity using fringe maxima/minima values $T_{\pm}$. The simplest approximation for the absorbance is $\alpha d = -\ln T_+$ is incorrect in all cases [16] [17]. Substituting $\Theta = \pi$ for maximum fringe amplitude into Eqn. 1 produces:

6. $\quad -\ln T_+ = \alpha d - \ln\left(\frac{\frac{(n^2+k^2)}{n^2}(1-R)^2}{1+R^2 e^{-2\alpha d} - 2Re^{-\alpha d}}\right)$

and the second term is not always negligible. We plot the error arising from ignoring it in Figure 3 across the entire $n - k$ parameter space relevant to this paper, and we find that the absorbance can be estimated incorrectly by up to a factor of two. In fact, the problem becomes more significant at high refractive indices and small absorption because the high-order internal reflections become more important to the total transmission of the sample – these multiple internal reflections undergo many passes of absorption; thus, the fringe amplitudes are more strongly attenuated than one would expect from the single-pass case.

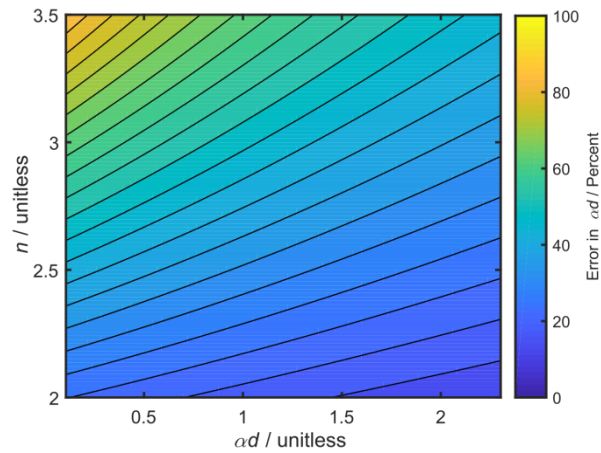

*Figure 3 – Systematic errors in the extracted absorbance-thickness product by assuming the fringe heights $T_{\pm}$ are described by the Beer-Lambert law, plotted for all combinations of $n$ & $k$ across the parameter space relevant to this paper. Errors peak at high values of $n$ and low $k$, where multiple internal reflections contribute most strongly to the shape of the transmission function.*

It is possible to rearrange Eqn. 6 to extract the absorption coefficient from $T_+$. It is also possible to extract it from the minima, $T_-$, or the "median" of the transmission, $T_0$. Substituting $\cos\Theta = b = +1, -1,$ or 0 into Eqn. 1 with corresponding transmission $T_b$, produces [16]:

7. $\quad e^{-\alpha d} = \dfrac{-(b \cdot 2T_b R - (1-R)^2) \pm \sqrt{(b \cdot 2T_b R - (1-R)^2)^2 - 4T_b{}^2 R^2}}{2T_b R^2}$

We have dropped terms in $k/n$, which we assume to be negligible since for typical samples in the terahertz region, $k \ll n$. Extracting the absorbance from data using Eqn. 7 is then possible as a simple extension to the fringe-finding method, as long as the positions $T_{\pm b}$ can be found from the data.

A few features are common to all variations of the fringe method. Obviously they are conditional upon accurately obtaining the values $T_b$ and $f_b$ from the data. Fringe methods are therefore strongly limited by discretization errors and random noise sources. Furthermore, the methods give the experimenter information solely about the refractive index at the specific points $f_b$; if we desire high resolution results we require optically thick samples for their closely spaced maxima and consequently a very high resolution FTIR instrument. Optically thick samples with high absorption coefficients tend to result in lower signal to noise ratio, exacerbating errors due to noise.

Aside from the fringe-differences technique, variations within this family of algorithms occur in the practical matter of obtaining the set of $T_b$ & $f_b$. A simple and fast way is to use a simple peak-finding algorithm, but any noise tends to require smoothing to be applied to the measured dataset. Fringe frequencies may be refined using either interpolation or a local fitting procedure around each fringe. Figure 4 compares a peak-finding algorithm without refinement to one with refinement via fitting, and we see that systematic errors due to discretization dominate peak-find algorithms, which refinement removes at the cost of significant computation time. We also observe in Figure 4 the aforementioned systematic error in the peak-differences method caused by the assumption that $n$ is not a function of frequency.

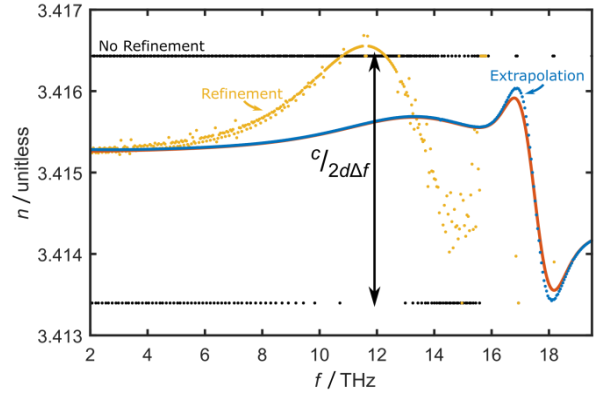

*Figure 4 – Refractive index extraction using different variations of the fringe-derived methods applied to the model data of Fig 1b. (black) The fringe-difference method without refinement of fringe location, showing the effects of discretisation error. (gold) The fringe-difference method with refinement, showing the systematic error due to variation in the refractive index. (red) The extrapolation method for obtaining the fringe order using refinement. (blue) Modelled real part of the refractive index, $n_{model}$.*

## *Fourier Methods & Hybrid Peak-Find Techniques*

Any function with periodic components is a candidate for analysis in its Fourier space, so we can use Fourier analysis methods to augment our algorithms for RI extraction. We now introduce the general working principles of Fourier analysis in this context, allowing us to find better refinement procedures for the peak-find methods, and in the next section to introduce a different variety of techniques which leverage the Fourier space more directly.

Consider two experimental interferograms; $s(\tau)$ the measured experimental signal through the sample, and $q(t)$ the comparable reference interferogram. Assuming that the reference measurement is made carefully, the two are related by a convolution with the inverse Fourier transform of the transmission function, $T(f)$:

8.  $s(\tau) = t(\tau) \otimes q(\tau) \leftrightarrow S(f) = T(f)Q(f)$

Where $\otimes$ denotes a convolution. $T(f)$ & $t(\tau)$, $S(f)$ & $s(\tau)$ and $Q(f)$ & $q(\tau)$ are Fourier transform pairs, and the right-hand side of Eqn. 8 follows directly from the Fourier convolution theorem.

If the experimental sample is a Fabry-Perot etalon, then $T(f)$ exhibits the characteristic fringes of Eqn. 1, which are induced by interference between the multiple internal reflections of the sample. Periodic features in reciprocal space, $T(f)$, have corresponding sharp features in the time domain, $t(\tau)$, appearing as a harmonic series leading away from the main centerburst as illustrated in Figure 5. We label these peaks with index $j$ so that $j = 0$ indicates the centreburst, $j = 1$ indicates the first order feature, etc. The precise shape of each feature encodes information about the variations in periodicity of $T(f)$ over the measured range.

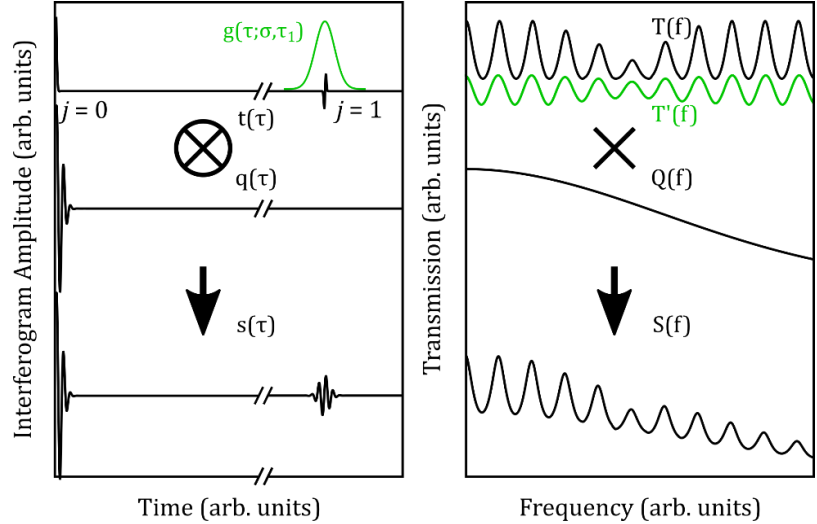

*Figure 5 – Formation of the interferogram by the transmission function. (a) the transmission function $t(\tau)$ is convolved with the reference interferogram $q(\tau)$ to form the additional structure in the measured interferogram $s(\tau)$. The Gaussian windowing function $g(\tau; \sigma, \tau_1)$ is shown in green. We have included a discontinuity on the time axis to indicate that the separation is much greather than the width of the features, which is the case if the fringe period in the frequency domain is much smaller than the variation in the fringes due to $\hat{n}(f)$ or the system response. (b) the transmission function $T(f)$ is multiplied by the reference spectrum $Q(f)$ to produce the total measured transmission spectrum $S(f)$. The inverse Fourier transform of $g(\tau; \sigma, \tau_1) \times t(\tau)$ is shown as $T'(f)$ in green.*

We suppose that the sample is sufficiently thin and that the refractive index is sufficiently slowly varying that the harmonics in $t(\tau)$ are well resolved. We multiply with a Gaussian window function in the time domain, $g(\tau; \sigma, \tau_1)$, where $\tau_1$ is the location in the time domain of the $1st$ order feature in $s(\tau)$, measured relative to the centreburst at $\tau_0$. The width parameter $\sigma$ is set such that the FWHM $\Delta\tau < \sigma\sqrt{2\ln 2} < \tau_1$, where $\Delta\tau$ is the width of the feature. The new function $t'(\tau; \sigma, \tau_1) = g(\tau; \sigma, \tau_1)t(\tau)$ is inverse Fourier transformed, whereby we obtain a complex-valued function $T'(f; \sigma, \tau_1)$ that has picked out the variation of the transmission function $T(f)$ which is oscillating with a well-defined periodicity:

9. $T'(f; \sigma, \tau_1) = G(f; \sigma, \tau_1) \otimes T(f)$

$G$ is the FT of $g$. Taking $T(f)$ to be an even function makes $s(\tau)$ real and even and allows us to take $g(\tau; \sigma, \tau_1)$ to be real and even, so that $G(f; \sigma, \tau_1)$ is real and even. Therefore $G$ is a cosinusoidally oscillating function with period equal to the average of the FP fringe periods and a Gaussian envelope. The width in frequency of $G$ is larger than the FP fringe period (since $g$ has a temporal width that is smaller than the separation between the harmonic burst features) and smaller than the frequency scale over which the fringe period varies (since $g$ has a temporal width that is larger than the width of the burst feature). The new function $T'(f)$ therefore retains the FP fringes of $T(f)$ and their variation in period, but eliminates their characteristic asymmetrical shape, along with the average background. The Gaussian windowing process has an additional advantage of rejecting much of the noise in the data, thus making $T'(f)$ more amenable to peak-finding methods than the raw data. This means it is easier to find $m$ and hence infer the real part of the refractive index.

Note that centring $g$ on the $j^{th}$ order feature instead of the $1^{st}$ means that peaks in $T'(f)$ have $j$ times higher density (and $n = \pi c/2\pi j d\Delta f$), and the resulting refractive index results are more finely resolved, at a cost of smaller signal since the 1st order in the FT of Eqn. 1 is the largest.

However, since the amplitude of $T'(f)$ is related in quite a complicated way to $k$, it is not accurate to use the processed fringe amplitudes to infer $k$ using Eqn. 7. A more advanced mathematical analysis may provide a good method for this, but it is outside the scope for this paper. Instead, returning to the raw data $T(f)$, the fringe amplitudes of which may be refined with fewer fitting variables now that $f_\pm$ are known from $T'(f)$. Figure 6 (a) compares $T(f)$ to $T'(f)$ and the extracted RI for each using the fringe-derived method with refinement by fitting; we see how a simpler function $T'(f)$ increases the contrast between maxima and minima in regions where the absorption is large, thus improving the accuracy of fringe location refinement efforts in the affected frequency band. The corresponding increase in accuracy of the recovered $n$ is demonstrated in Figure 6 (b).

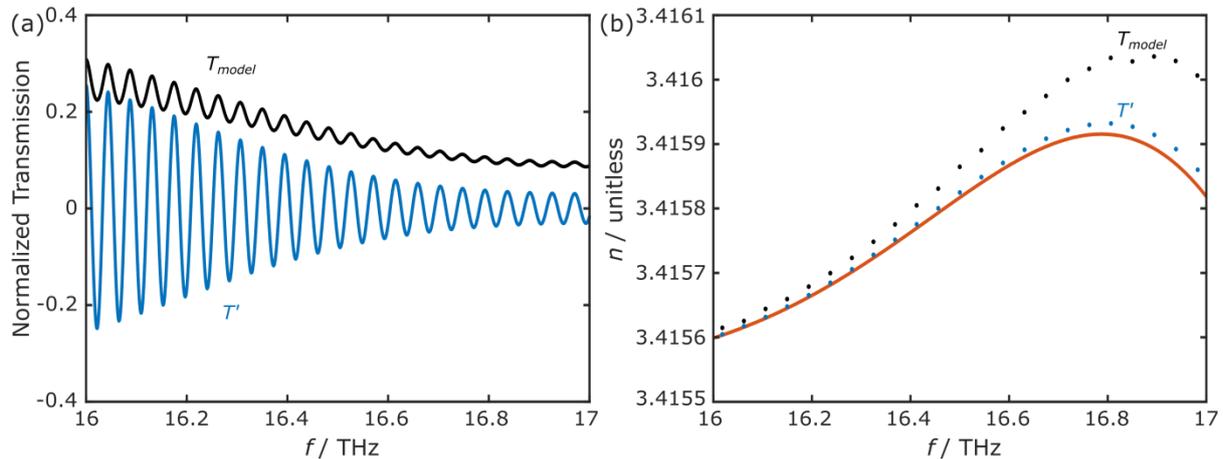

*Figure 6 – Effect of windowing the transmission function. (a) The model data $T_{model}$ (black) is compared to the transmission function windowed at $j = 1$, $T'$ (normalized), showing how the oscillatory nature of the function is retained, and the relative fringe amplitudes enhanced by windowing in the Fourier domain. (b) Effect of this increased contrast on the extracted $n$ near the main transient in $n_{model}$, showing the dramatic increase in accuracy.*

While processing in the time domain has been studied in the literature, available manuscripts choose to inverse Fourier transform a subset of the data near the 1st harmonic peak rather than apply a smooth window, as we have presented here [23] [19] [24]. We believe that these choices were made because of the challenges of computation when the methods were first derived, rather than for any inherent advantage. The mathematics of smooth apodising functions used in this work is widely appreciated, and peak refinement techniques have been applied separately [10]. Nevertheless, to the knowledge of the authors, the combination of peak refinement with Fourier windowing described in this section has not been previously reported in the literature. We propose that Fourier filtering can be adopted as a simple additional processing step which will increase the accuracy of RI extraction for any experimenter who uses a fringe-derived method

### Phase Methods

Let us now assume that our measured dataset has unknown variations in the refractive index, so an extrapolation method might be vulnerable to systematic errors. We would like to have an alternative tool available that may be used as a cross-comparison and that would suffer from systematic errors under different conditions. A more sophisticated analysis also might hope to obtain higher datapoint densities, perform better using thinner samples, and still be robust to noise. Here we make use of the Fourier analysis of the last section to introduce a fundamentally different family of methods which meet these expectations.

A general approach can be defined by writing out the derivative of the internal phase factor of the sinusoidal term of Eqn. 1 with respect to frequency, now explicitly acknowledging that $n$ is a function of $f$. The numerical derivative of $\Theta$ with respect to $f$ is:

10. $\frac{\Delta\Theta}{\Delta f} = 2\frac{\Delta\phi}{\Delta f} + 2\frac{2\pi}{c}nd + 2\frac{2\pi}{c}fd\frac{\Delta n}{\Delta f}$

Let us assume that the phase change upon reflection is entirely negligible (an assumption we shall justify later):

11. $\Delta\Theta = 2\frac{2\pi}{c}d(n \cdot \Delta f + f \cdot \Delta n)$

A measurement of the fringe period $\Delta f|_{\Delta\Theta=2\pi}$ is enough to extract n so long as $\Delta n$ is negligible (i.e. $dn/df$ is negligible) in which case we recover the fringe-difference method from Eqn. 5. When the RI varies strongly and $dn/df$ is non-negligible, systematic errors appear in the fringe-difference method as discussed earlier. We require a relationship between the change in phase of the transmission function $\Delta\Theta$ over any given frequency interval $\Delta f$, and the refractive index.

Several algorithms have been demonstrated that extract information from the Fourier space [24] [23] [19] [20], which obtain the phase change through manipulation of $T(t)$. Such "Fourier phase" algorithms yield datapoint densities similar to that of the input data, thus allowing the rate of change of the refractive index to be high and measured at every datapoint in the source data. Fourier methods therefore do not require thicker samples to obtain dense sampling, and so thinner samples may be measured – this in turn allows materials with higher absorption indexes to be analysed. Most notably, they allow FTIR instruments with lower frequency resolution to be used.

Present methods still require additional information so that Eqn. 11 may be solved, which has caused these implementations to resort to extrapolation of $\Theta$ to $f = 0$. We thus encounter the same basic problem as fringe methods, and require a new approach. We propose that the problem be addressed using a single global free parameter, by adjustment of which we may obtain the entire spectrum of $n(f)$. If Eqn. 11 is discretised such that a sample of index $i$ has a measured change in phase $\Delta\Theta_i(f_i)$, then we may write a recurrence relation:

12. $\Delta\Theta_i = 2\frac{2\pi}{c}d\big(n_{i-1} \cdot \Delta f + f_i \cdot (n_i - n_{i-1})\big)$

Here, $\Delta f$ is the point spacing. Now if the refractive index at any fixed frequency along the interval, $n_0(f_0)$, is known and $\Delta\Theta_i$ can be experimentally determined, then the entire spectrum of $n_i$ may be found from the relative phase change from point to point, without needing to know the absolute phase. A judicious choice of $f_0$ & $n_0$ is therefore required, but this can be estimated from the spacing of the features in $T(t)$ and further refined by studying the systematic errors caused by incorrect guesses for $n_0$. By avoiding extrapolation we have found a substantially different way to express the problem.

The remaining step in the algorithm is to obtain $\Delta\Theta_i(f_i)$ from the experimentally measured $T(f_i)$ by using almost the same windowing method as previously Eqn 9, except that we do not make $g(\tau;\sigma,\tau_1)$ even, we use a single Gaussian only at positive $\tau_1$, so that $G(f;\sigma,\tau_1)$ is the same Gaussian envelope but modulated with $e^{-if(\tau_1-\tau_0)}$ instead of a real cosine function. This preserves the phase information in the argument of the result, which may be written.

13. $T'(f) = A(f)e^{if(\Theta_0+\delta\Theta)}$

where $\delta\Theta$ is the relative phase, which is a function of frequency due to variation in $n$, and $\Theta_0$ is an (irrelevant) constant that is determined by both the (unknown) absolute phase and the (arbitrary) phase choice for $g(\tau;\sigma,\tau_1)$. $A(f)$ is an amplitude that is real and slowly varying so long as $R, k, \phi$ are slowly varying. $A(f)$ is also irrelevant since, as mentioned above it depends on k in a complicated way. Effectively, the windowing has demodulated the quickly varying FP fringes by a factor of $t_1$ to recover only a slow variation due to the change in refractive index from its baseline. We therefore obtain $\delta\Theta(f)$ by finding the argument of the complex valued $T'(f)$ Eqns. 9 & 13. Once discontinuities in the phase have been removed [27], $\Delta\theta_i(f_i)$ is found numerically. Note that $\tau_1$ is determined relative to $\tau_0$, the centreburst position, and potential systematic errors in estimating $\tau_0$ may be removed by repeating the procedure to find $\delta\Theta(f)$ using $g(\tau; \sigma, -\tau_1)$ and taking the mean. As before, higher orders may be used but they are weaker.

Systematic errors inherent to the algorithm stem from two primary features. First, since the algorithm is a recurrence relation, any error in $n_{i-1}$ extends to some degree to the extracted $n_i$. Assuming an error $\varepsilon$ is introduced at index $i^{(\varepsilon)}$, its influence is propagated to following indices in a manner which converges back to the correct solution as $i > i^{(\varepsilon)}$, as supported theoretically in Appendix 1. Consequently, an error in the estimate of $n_0$ has a strong influence upon systematic errors in the extracted RI, as shown in Figure 7. Errors in $n_{i-1}$ may occur due to an extracted $\delta\Theta$ not representing the actual change in RI, for example if $R, k, \phi$ have strong periodic components through some part of the dataset. Second, since the windowing procedure is equivalent to a convolution, the measured $\delta\Theta$ at each end of the dataset will be smoothly pinned to zero as shown in Figure 7. This error may be mitigated by discarding the relevant datapoints between $f_{min}$ & $f_{min} + 1/(\sigma\sqrt{2})$ prior to applying the recurrence relation. Major deviations from the expected transmission function can also cause this kind of error, for example where background calibrations become inaccurate due to low source intensity or beam splitter transmission edges etc.

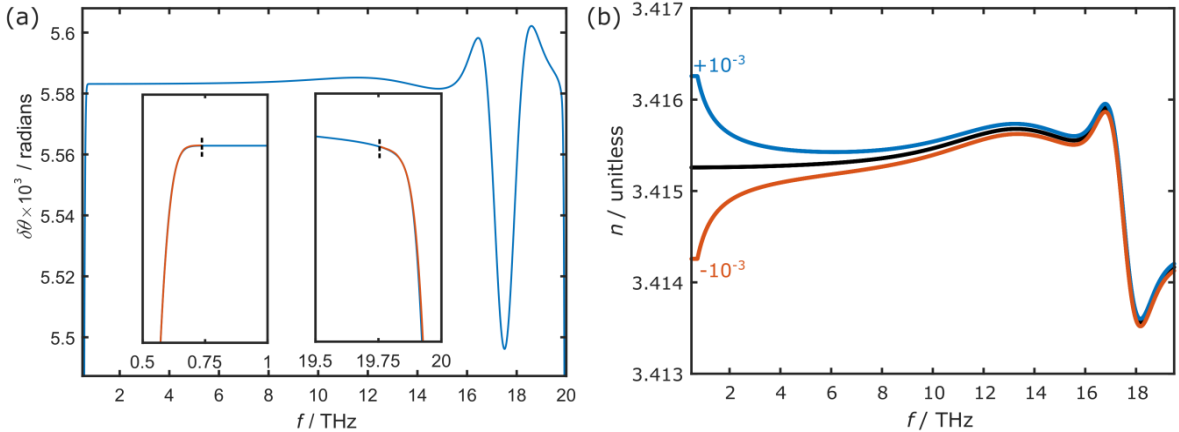

*Figure 7 – Systematic errors in the phase method. (a) extracted phase change $\delta\theta$ from example data as a function of frequency $f$, showing how the convolution effect dramatically suppresses the phase change near the limits of the dataset. Inset: views of the extracted phase change on a magnified frequency axis, with the cutoff regions shown in red. (b) Systematic errors in extracted RI caused by incorrect initial guess $n_0$; (black) correct guess replicating the input very well, (blue) incorrect guess overestimating $n_0$ by 0.001, (red) incorrect guess underestimating $n_0$ by 0.001. Flat regions near $f = 0$ are regions shown inset in (a).*

We have now a complete process for obtaining $n$, but the extraction of $k$ is not so straightforwardly obtained. The local average of the function $T(f)$ over one period of the oscillation, which we will denote $T_A(f)$, is analogous to a standard integral:

14. $T_A(f) = \frac{c_3}{2\pi} \int_{c_3 f' = 2\pi(m-1)}^{c_3 f' = 2\pi m} \frac{c_0 df'}{c_1 + c_2 \cos(c_3 f')} = \frac{c_0}{\sqrt{c_1{}^2 - c_2{}^2}}$

where the $c_{0,1,2,3}$ are defined by setting the integrand equal to Eqn. 1 and neglecting variations in $c_{0,1,2,3}$ over each period. Hence:

15. $T_A(f) = \frac{(1-R)^2 \, e^{-\alpha d}}{1 - R^2 e^{-2\alpha d}}$

We have, as before, dropped terms in $k/n$ and assumed that the terms in $R$ are sufficiently slowly varying for an adequate approximation. Eqn. 15 has been obtained in the literature by summation of intensities [17] [28], and can also be obtained by setting $\gamma = 0$ in Eqn. 1, whereas the presented analysis gives a straightforward argument for the same result based on the coherent properties of Eqn. 1 with $\gamma = 1$. Since the local average of $T(f)$ is defined by the centerburst of the interferogram, we may find $T_A(f)$ experimentally by applying the FT window $g(\tau; \sigma, \tau_0)$. Importantly, this approach also yields results that are entirely independent of coherence of the experiment. Since $\gamma$ generally has its own spectrum independent of the parameters we are interested in, we have proposed a method which may entirely avoid systematic errors due to an imperfect experiment or sample in a way only achievable previously using laborious fitting techniques [22]. Although extracting $\gamma(f)$ is outside the scope of this work, our result implies that a high resolution measurement of the coherence fraction can be obtained and used to study spectrometer resolution or surface roughness.

As we required earlier, we can now extract both the real and imaginary parts of the refractive index using the same methodology – the total algorithm is shown diagrammatically in Appendix 2. We now use our model data to demonstrate that our algorithm can recover known input data from a model transmission function. We find that the recovered $\hat{n}$ is sufficiently accurate to be essentially indistinguishable from the model inputs on the scales of Figure 1, so we show the relative error in the extracted refractive index $\Delta n/n_{model}$ expressed in parts per million (ppm) in Figure 8 (a). Systematic errors in the refractive index are consistently small across the spectrum, and they do not correlate with regions where we dropped terms in Eqn. 10 due to $\phi$ (the relative weight of which is $\Delta\phi/\Delta\Theta$, which we analyse mathematically in Appendix 3) dominate over those terms we retained, shown on the right axis of Figure 8 (a).

This potential systematic error is not strongly propagated along the frequency axis despite the iterative nature of the algorithm due to the asymmetric nature of $\Delta\phi/\Delta\Theta$ around its poles. The observable gross-scale systematic errors are caused by the finite width nature of the Gaussian $g(\tau; \sigma, \tau_j)$, which inevitably induces a degree of smoothing of the complex number $e^{if\delta\Theta}$ in proportion to the reciprocal of its width. This is in turn limited by the separation of the maxima $t_{j+1} - t_j$, which can be controlled by the thickness of the sample $d$; systematic errors may thus be reduced using this technique by *reducing* the thickness of the sample, although the degree to which this is possible will be limited by uncertainties in the thickness of very thin samples.

We then turn our attention to the extraction of $k(f)$, and find that the extracted imaginary part is also almost indistinguishable from the defined $k_{model}$. The absolute error in $k$ is shown in Figure 8 (d) (right axis), normalized to the peak value of $k_{model}$ (relative errors diverge because $k_{model} \to 0$ in large parts of the spectrum). We compare the errors in extraction with errors in the algorithm's estimate of $T_A(f)$ (left axis), which quantifies the error due to the approximation that the terms in $R$ of Eqn. 14 are slowly varying. Errors due to this approximation are small in this case, but are systematic and will grow as the refractive index varies more rapidly.

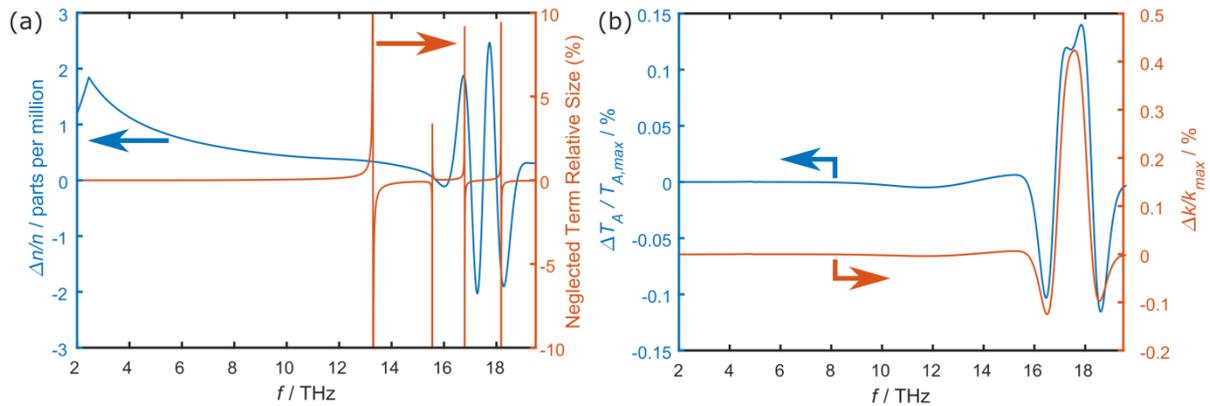

*Figure 8 – Model-based testing of refractive index extraction algorithms. (a) (left axis) relative error in extracted $n$, in units of parts per million; (right axis) relative weight of the neglected term of Eqn. 10 due to phase change upon reflection, compared to the term kept for use in the phase method, expressed as a percentage. (b) (left axis) absolute error in measurement of the incoherent transmission function $T_A(f)$ by the phase method normalized to the maximum value of the incoherent transmission function, expressed as a percentage; (right axis) absolute error in extracted $k$ using the phase method, normalized to the maximum value of $k$, expressed as a percentage.*

## Results

Our study now proceeds to analyse the relative performance of optimized fringe-fitting methods compared to our proposed Fourier phase method in cases where experimental non-idealities are present. We start by modelling these experimental limitations using simple extensions to our model data presented in previous sections, and thereafter apply the algorithms to real experimental data for a practical comparison. By doing so, we gain an insight into the real limitations on refractive index extraction in contemporary settings comparable to travelling standard analysis [10].

### Testing Algorithms

Canonical studies focus on the practical extraction of refractive index from a real dataset, but algorithms' inherent flaws should also be studied systematically. Metrological studies generally require a controlled test which is sufficiently independent of experimental limitations, especially here where the precise values of $\hat{n}$ are not already well studied. To demonstrate this principle, we compare our new algorithm to a robust peak-find method which uses Fourier windowing, refinement by fitting, and extrapolation techniques discussed in the Theory section. The two algorithms and their input parameters were optimized in isolation on sample $T_{model}$ datasets without simulated experimental non-idealities.

To measure the comparative robustness of different algorithms, we choose the root mean square (RMS) error as a figure of merit. Better performing algorithms will show less overall deviation from the correct result, and this should be reflected in the RMS error – lower is better. Random noise and low resolution are simulated and added to $T_{model}$ separately. For simplicity, we use a multiplicative factor of $(1 + r)$, where $r$ is a random number drawn from a normal distribution of controlled width. By varying the width of the distribution, we vary the noise amplitude in the model dataset. Errors due to low resolution sometimes termed "slit-width" errors [16], where the spectrometer's resolution is insufficient to perfectly resolve $T_{model}$, are simulated by applying a linear averaging function of varying width.

Figure 9 shows that our Fourier phase method outperforms our implementation of the fringe method under almost all circumstances, apart from cases where the input data are essentially perfect. The RMS error in $n$ from the phase method is always lower in our example, although this will always be limited by whichever method happens to have the smaller instances of systematic error. If the algorithms show similar robustness to noise, one must be careful not to overvalue relative differences in errors which might depend on details of the implementation.

Rather, a difference in the scaling of a quantity would be the most persuasive argument for any competitive algorithm's superiority. Figure 9 shows that the fringe method's RMS error in $k$ increases dramatically as the experimental non-idealities scale up, whereas the phase method's RMS error in $k$ remains constant. This is true both in the case of random noise, Figure 9 (a), and in the case of resolution errors, Figure 9 (b). These kinds of non-idealities represent a range of experimental situations which can be difficult to minimize – detector sensitivity and noise floor, step size and repeatability, and so on. The Fourier phase method is insensitive to these errors since it deliberately averages $T$ as part of the analysis, which is a major reason to choose it over a fringe method.

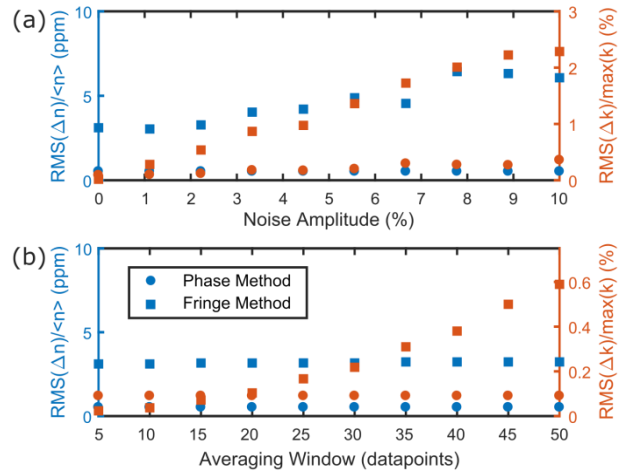

*Figure 9 – Comparative robustness of the Fourier method algorithm. Relative error in $n$ (blue, left axis) and $k$ (red, right axis) extracted using the Fourier method (circles) and an optimized fringe method (squares) for the refractive index model of Figure 1 under two different noise models. (a) random noise simulate by a multiplicative Gaussian distributed noise of controlled amplitude; (b) finite-slit-width error simulated by a finite-size local smoothing of a controlled window width, where 50 datapoints equates to roughly $1/10^{th}$ of a fringe near $f = 10\ THz$.*

### Application to Silicon Optical Flats

We demonstrate the Fourier algorithm's practical application by measuring the transmission function of a sample of high resistivity float-zone silicon at room temperature. The wafers were purchased commercially with a thickness of $d = (1.0 \pm 0.1)\ mm$, and measured to be of thickness $d = (1.070 \pm 0.005)\ mm$ using profilometry. The transmission spectrum was measured using a Bruker IFS125HR with a datapoint separation of $0.004\ cm^{-1}$ ($0.1\ GHz$) and a nominal resolution of $0.01\ cm^{-1}$ ($0.3\ GHz$) over the region $f = 30 - 660\ cm^{-1}$ ($0.9 - 19.8\ THz$). Our measurement of the transmission function $T_{Si}(f)$ is shown in Figure 10 (a), which is qualitatively similar to the modelled transmission function of Figure 1 – Model data used in the theoretical parts of this work. (a) Modelled complex refractive index derived by numerical KK transform; real part (blue) derived from the defined imaginary part (red). (b) Model transmission function $T_{model}$ obtained by substituting the data from (a) into Eqn. 1 with an etalon thickness of $d = 1\ mm$ and $\gamma = 1$.Figure 1 (b). The algorithm was applied with an initial guess $n_0 = 3.4157$, and we show the extracted $\hat{n}$ in Figure 10 (b). Figure 10 (b) also shows the same analysis using our implemented fringe method discussed above. We find an absorption in the region of the transverse-optical (TO) phonon

c. $18.5\,THz$ [22] [15] and a corresponding rapid variation in the refractive index. Similarly, a long-baseline variation in the absorption due to the phonon band is matched by a consistent variation in the real part of the refractive index. Both algorithms' extracted $\hat{n}$ compare well to historical studies of Si [22] [17] [15], with improved datapoint density, but historical comparisons are of limited value where the same test artefact is not used. Both methods' extracted parameters agree around the extrema of the dataset, particularly at the TO phonon absorption peak. However, there is a major disagreement around $12\,THz$ where the phase method extracts only a negligible absorbance and the fringe method extracts a significant value.

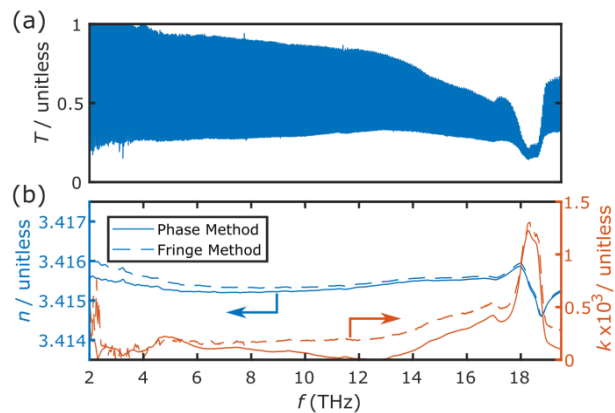

*Figure 10 – Application to a Si optical flat similar to a metrological travelling standard. (a) FTIR transmission spectrum measured at high resolution; (b) extracted $n$ & $k$ using the phase algorithm (solid lines) using an optimized fringe method (dashed lines).*

We can probe this further by substituting our extracted parameters back into Eqn. 1 to recreate the transmission function. We find that using the $n$ and $k$ extracted from the phase method dramatically overestimates the sharpness of the FP fringe maxima near $12\,THz$, and gives the impression of a poor agreement, whereas the fringe method gives larger $\alpha$ with broader fringes, that superficially look in better agreement. This discrepancy is due to the fact that the fringe method extraction of $n$ and $k$ (and $\alpha$) is sensitive to $\gamma$ (see Appendix 4) but the phase method is not (Eqn. 15). The fringe method cannot be used to extract $\alpha$ and $\gamma$ independently, and some assumption must be made. Our implementation takes the approximation that $\gamma = 1$ (negligible incoherence) and has produced a high $\alpha$ around $12\,THz$ in order to match the data, when in fact $\alpha$ drops below the level at which that the effect of $\gamma \neq 1$ starts to become noticeable. The effect of $\gamma \neq 1$ in the fringe method thus produces a complicated systematic error, especially important when $\alpha$ is small, whereas the phase method produces a systematically better measurement of $k$. Clearly it is dangerous to look at the superficial agreement in the frequency domain, or to extract information from the absolute fringe amplitude/sharpness.

Our experimental observations therefore underline the ability of the new phase method to discriminate absorption effects from other loss processes such as scattering or low resolution, without requiring both transmission and reflection measurements. Although extracting $\gamma(f)$ is beyond the scope of this paper, we anticipate applications of this feature of the algorithm to study imperfections in samples over large surfaces or volumes. All of our data and analysis techniques are published for free under the GNU general license [29].

## Conclusion

The metrological applications of FTIR, and its relevance to the challenging regions of the terahertz spectrum, motivate us to explore new metrological techniques that take advantage of a common and well understood instrument. We have presented a review of some methodologies for extracting refractive indices from FTIR measurements of optically flat samples, and used the overview as an opportunity to suggest a new technique that makes the most of the computational ability to filter in the Fourier space. We showed that it is possible to quantify the systematic errors in the algorithm by applying it to model data, and in doing so have demonstrated a thorough metrological analysis of these algorithms. This kind of methodology can allow future intercomparisons between algorithms using a standardised test which allows for comparison to exact analytic results. To ensure that fair comparisons can be drawn and our conclusions replicated, our analysis tools, data, and code for each technique discussed in the paper have been published under the GNU general license [29].

Our proposed algorithm yields the complex refractive index with a very fine datapoint spacing on the order of that of the input data, a major benefit over more commonly utilized fringe methods. We have also discussed how our method prefers to analyse optically thin samples, whereas fringe methods would prefer thick samples – this enables us to measure the optical constants of the material accurately in the presence of higher absorption and with lower frequency resolution. The robustness of the method to incoherence is significant for metrology, since surface roughness and volume inhomogeneities in physical artefacts can be ignored as long as a good measurement of the mean thickness of the samples is obtained.

By the same token, our analysis has also for the first time dedicated significant effort to analysing the limitations of the algorithm, which is a crucial contribution for advancing metrology in the terahertz region. Finally, our technique has managed to avoid the inherent difficulties of KK transformation as part of the extraction technique, so that we avoid some sources of systematic error in other analyses [17]. It is also proposed that some of the limitations we have discussed in this paper can be mitigated by using the same principles we have outlined to form an iterative improvement procedure, much like that of King *et al* [17]. Scattering terms might also be incorporated into the model of Eqn. 1 so that inferences can be drawn about the non-ideality of the samples under test without requiring additional reflectance measurements.

While our algorithm has significant advantages over others, we have established the different systematic errors in each of the different algorithms, and so none can be considered universally applicable. A judicious choice must be made based on the shape of the transmission function, the resolution of the spectrometer, and the approximate optical thickness of the sample. Our new algorithm is therefore part of a toolbox of different techniques which would allow a metrologist to select the best analysis for a given dataset.

In cases where one wishes to be especially rigorous, we envision that an experimenter will first obtain a transmission spectrum of the material of interest, from which an estimate of the absorption spectrum can be obtained. Using the KK transform, the experimenter can then obtain an estimate of the refractive index variation in the sample and test a variety of extraction algorithms on model data, so that the best technique may be used on the real data. An intercomparison using the RMS error from each technique might allow for quantitative optimization of the analysis, and thus a more robust metrological process to support intercomparison efforts and calibration standards. Automation of this procedure could even allow the choice of the best analysis algorithm without user intervention, thus making the best measurement of complex refractive index straightforward even for a non-expert user.

## Acknowledgements

We acknowledge financial support from the UK Engineering and Physical Sciences Research Council [COMPASSS/ADDRFSS, Grant No. EP/M009564/1].

The central data for our paper and our code analysis tools are available to the public from a permanent data repository [29].

# Appendix 1: Propagation of Errors in the Fourier Phase Algorithm

Starting from the recurrence relation:

$$\Delta\Theta_i = 2\frac{2\pi}{c}d\big(n_{i-1}\cdot\Delta f + f_i\cdot(n_i - n_{i-1})\big)$$

We rearrange to find $n_i$:

$$n_i = \frac{1}{f_i}\frac{c}{4\pi d}\Delta\Theta_i + n_{i-1}\frac{f_{i-1}}{f_i}$$

By iterative substitution we can find any $n_i$ in terms of its previous iterand:

$$n_{i-1} = \frac{1}{f_{i-1}} \frac{c}{4\pi d} \Delta\Theta_{i-1} + n_{i-2} \frac{f_{i-2}}{f_{i-1}}$$

Therefore the ith iteration is in terms of the i-2th iteration:

$$n_i = \frac{1}{f_i} \frac{c}{4\pi d} \Delta\Theta_i + \frac{1}{f_i} \frac{c}{4\pi d} \Delta\Theta_{i-1} + n_{i-2} \frac{f_{i-2}}{f_i}$$

If we deliberately introduce an error by substituting $n_{i-2} \rightarrow n_{i-2} + \varepsilon_{i-2}$:

$$n_i + \varepsilon_i = \frac{1}{f_i} \frac{c}{4\pi d} \Delta\Theta_i + \frac{1}{f_i} \frac{c}{4\pi d} \Delta\Theta_{i-1} + n_{i-2} \frac{f_{i-2}}{f_i} + \varepsilon_{i-2} \frac{f_{i-2}}{f_i}$$

The final term is the error in $n_i$, which we can generalise to any error introduced $I$ iterations previously:

$$\varepsilon_i = \frac{f_{i-I}}{f_i} \varepsilon_{i-I}$$

Any error introduced will always reduce in contribution to the total error, by a scaling rule that looks like $1/f$ as shown experimentally in the main text. What if the error is instead in the phase, which we argue in the paper is the cause of most of our systematic errors? Recall:

$$n_i = \frac{1}{f_i} \frac{c}{4\pi d} \Delta\Theta_i + \frac{1}{f_i} \frac{c}{4\pi d} \Delta\Theta_{i-1} + n_{i-2} \frac{f_{i-2}}{f_i}$$

If we make the replacement $\Delta\Theta_{i-1} \rightarrow \Delta\Theta_{i-1} + \epsilon_{i-1}$:

$$n_i + \varepsilon_i = \frac{1}{f_i} \frac{c}{4\pi d} \Delta\Theta_i + \frac{1}{f_i} \frac{c}{4\pi d} \Delta\Theta_{i-1} + n_{i-2} \frac{f_{i-2}}{f_i} + \frac{1}{f_i} \frac{c}{4\pi d} \epsilon_{i-1}$$

This is reminiscent of the case above, so we can infer that for any phase error $\epsilon_{i-I}$:

$$\varepsilon_i = \frac{c}{4\pi d} \frac{1}{f_i} \epsilon_{i-I}$$

Phase errors therefore drop in significance in a similar manner to errors in $n$.

## Appendix 2: Implementing the Fourier Phase Algorithm

Here we represent the phase algorithm for extracting the complex refractive index in a flow diagram, showing the main calculation and decision steps:

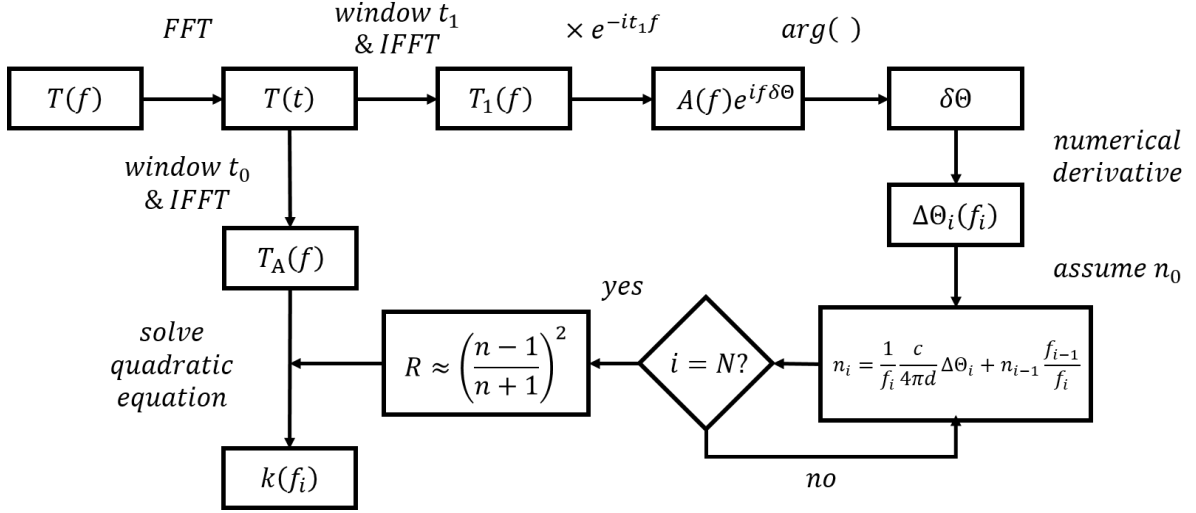

## Appendix 3: Systematic Errors Due to Dropped Terms

In our main analysis, we make the assumption that the rate of change of the phase shift upon reflection is negligible. To study mathematically whether this is the case and how robust our algorithm is if this assumption is dropped, we split the derivative into two components:

$$\frac{d\Theta}{df} = \frac{d}{df}\left[2\phi + 2nf\frac{2\pi}{c}d\right] = A + B$$

We assumed $A = 0$ to derive the algorithms in the phase method. Since:

$$\tan\phi = \frac{2k}{n^2 + k^2 - 1}$$

We will naturally find that $A$ is quite a complicated quantity. We can make the assumption that $k \ll n$ and apply an approximation to find:

$$A(k \ll n) = \frac{4}{(n^2 - 1)^2}\left((n^2 - 1)\frac{dk}{df} + 2kn\frac{dn}{df}\right)$$

Hence by substitution:

$$\frac{d\Theta}{df} = \frac{4}{(n^2 - 1)^2}\left((n^2 - 1)\frac{dk}{df} + 2kn\frac{dn}{df}\right) + 2\frac{2\pi}{c}d\left[n + f\frac{dn}{df}\right]$$

In the regime where $f \sim 10^{12}\ Hz$ then it is reasonable to conclude that:

$$\frac{4}{(n^2 - 1)^2}2kn\frac{dn}{df} \ll 2\frac{2\pi}{c}df\frac{dn}{df}$$

So we can drop the term on the LHS above and arrive at the estimate:

$$\frac{d\Theta}{df} = \frac{4}{(n^2 - 1)}\frac{dk}{df} + 2\frac{2\pi}{c}d\left[n + f\frac{dn}{df}\right]$$

The algorithm will tend to be most wrong near stationary points in $n$, which happen to correlate with peaks in $\frac{dk}{df}$. If $\frac{dn}{df}$ dominates the term in square brackets above, then the relative error in $\Delta\Theta$ will be of opposite sign around each maximum, as observed in the Theory section in the main work. Since the sign of the phase error changes around the maximum, error propagation through the spectrum is minimized by compensation through the mechanism discussed in Appendix 1.

## Appendix 4: Systematic Errors Due to the Coherence Factor

Here we study the effect of the coherence fraction upon the peak extraction method. In the perfectly coherent case $\gamma = 1$:

$$e^{-\alpha d} = \frac{-(b \cdot 2T_{\pm b}R - (1-R)^2) \pm \sqrt{(b \cdot 2T_{\pm b}R - (1-R)^2)^2 - 4T_{\pm b}{}^2 R^2}}{2T_{\pm b}R^2}$$

However, in the general case return to Eqn. 1 of the main work that:

$$\frac{\frac{1-\gamma^2 e^{-2\alpha d}R^2}{1-e^{-2\alpha d}R^2} \frac{(n^2+k^2)}{n^2}(1-R)^2 \, e^{-\alpha d}}{1+\gamma^2 R^2 e^{-2\alpha d} - 2\gamma R e^{-\alpha d}\cos[\Theta]}$$

By making the same replacement for $\cos[\Theta]$:

$$T_{\pm b} = \frac{1-\gamma^2 e^{-2\alpha d}R^2}{1-e^{-2\alpha d}R^2} \; \frac{\frac{(n^2+k^2)}{n^2}(1-R)^2 \, e^{-\alpha d}}{1+\gamma^2 R^2 e^{-2\alpha d} - 2b\gamma R e^{-\alpha d}}$$

If we let:

$$x = e^{-\alpha d}$$

Then the equation can be expressed as:

$$T_{\pm b}(1-x^2 R^2)(1+\gamma^2 R^2 x^2 - 2b\gamma Rx) = (1-\gamma^2 x^2 R^2)\,(1-R)^2 \, x$$

At this point it is clear that we must solve a quartic equation which does not have the convenience of becoming a simple quadratic in $y$ under the substitution $y = x^2$. While the equation must have solutions for $x$ in terms of $\gamma, R, b$ it should not be considered to reduce to the quadratic solution given in the main work, hence the peak method fails. The quartic equation does have a general solution, which is vastly more complicated than we consider within the scope of this paper. Either way, it is reasonable to suppose that an independent estimate for $\gamma$ is essential to extract $\alpha$, which is not acceptable from the position of requiring minimal a-priori assumptions about the sample's properties.